\documentclass[journal]{IEEEtran}

\usepackage{amsmath,epsfig,dsfont}
\usepackage{subfigure}
\usepackage{algorithm}
\usepackage{algorithmic}
\usepackage{amssymb,bm}
\usepackage{amsfonts}
\usepackage{stfloats}
\usepackage{multirow}

\begin{document}
\title{Indian Buffet Game with Negative Network Externality and Non-Bayesian Social Learning}

\author{\authorblockN{Chunxiao~Jiang\authorrefmark{1}\authorrefmark{2}, Yan~Chen\authorrefmark{1}, Yang~Gao\authorrefmark{1}, and K. J. Ray~Liu\authorrefmark{1}} \\ 
      \small\authorblockA{\authorrefmark{1}Department of Electrical and Computer Engineering, University of Maryland, College Park, MD 20742, USA\\ 
          \authorrefmark{2}Department of Electronic Engineering, Tsinghua University, Beijing, 100084, P. R. China\\ 
        E-mail:\{jcx, yan, yanggao, kjrliu\}@umd.edu}}

\maketitle

\begin{abstract}
How users in a dynamic system perform learning and make decision become more and more important in numerous research fields. Although there are some works in the social learning literatures regarding how to construct belief on an uncertain system state, few study has been conducted on incorporating social learning with decision making. Moreover, users may have multiple concurrent decisions on different objects/resources and their decisions usually negatively influence each other's utility, which makes the problem even more challenging. In this paper, we propose an Indian Buffet Game to study how users in a dynamic system learn the uncertain system state and make multiple concurrent decisions by not only considering the current myopic utility, but also taking into account the influence of subsequent users' decisions. We analyze the proposed Indian Buffet Game under two different scenarios: customers request multiple dishes without budget constraint and with budget constraint. For both cases, we design recursive best response algorithms to find the subgame perfect Nash equilibrium for customers and characterize special properties of the Nash equilibrium profile under homogeneous setting. Moreover, we introduce a non-Bayesian social learning algorithm for customers to learn the system state, and theoretically prove its convergence. Finally, we conduct simulations to validate the effectiveness and efficiency of the proposed algorithms.
\end{abstract}


\newtheorem{proposition}{Proposition}
\newtheorem{definition}{Definition}
\newtheorem{theorem}{Theorem}
\newtheorem{corollary}{Corollary}
\newtheorem{lemma}{Lemma}
\section{Introduction}

In a dynamic system, users are usually confronted with uncertainty about the system state when making decisions. For example, in the field of wireless communications, when choosing channels to access, users may not know exactly the channel capacity and quality. Besides, users have to consider others' decisions since overwhelming users sharing a same channel will inevitably decrease everyone's average data rate and increase the end-to-end delay. Such a phenomenon is known as negative network externality \cite{externality}, i.e., the negative influence of other users' behaviors on one user's reward, due to which users tend to avoid making the same decisions with others to maximize their own utilities. Similar phenomenons can be found in our daily life such as selecting online cloud storage service and choosing WiFi access point. Therefore, how users in a dynamic system learn the system state and make best decisions by considering the influence of others' decisions becomes an important research issue in many fields \cite{Liu}-\cite{Yan}.

Although users in a dynamic system may only have limited knowledge about the uncertain system state, they can construct a probabilistic belief regarding the system state through social learning. In the social learning literatures \cite{sl1}\nocite{bayeslearning,sl2,nonbayes}-\cite{nonbayes1}, different kinds of learning rules were studied where the essential objective is to learn the true system state eventually. In most of these existing works, the learning problem is typically formulated as a dynamic game with incomplete information and the main focus is to study whether users can learn the true system state at the equilibria. However, all of them assumed that users' utilities are independent of each other and thus did not consider the concept of network externality, which is a common phenomenon in dynamic systems and can influence users' utilities and decision to a large extent.

To study the social learning problem with negative network externality, in our previous work \cite{tomky1}\nocite{biling}-\cite{tomky2},we have proposed a general framework called "Chinese Restaurant Game". The concept is originated from Chinese Restaurant Process \cite{crp}, which is used to model unknown distributions in the non-parametric learning methods in the field of machine learning. In the Chinese Restaurant Game, there are finite tables with different sizes and finite customers sequentially requesting tables for meal. Since customers do not know the exact size of each table, they have to learn the table sizes according to some external information. Moreover, when requesting one table, each customer should take into account the subsequent customers' decisions due to the limited dining space in each table, i.e., the negative network externality. Then, the Chinese Restaurant Game is extended to a dynamic population setting in \cite{infocom}, where customers arrive at and leave the restaurant with a Poisson process. With the general Chinese Restaurant Game theoretic framework, we are able to analyze the social learning and strategic decision making of rational users in a network.

One underlying assumption in the Chinese Restaurant Game is that each customer can only choose one table. However, in many real applications, users can have multiple concurrent selections, e.g., mobile terminals can access multiple channels and users can have multiple cloud storage services. To tackle this challenge, in this paper, we propose a new game, called \textbf{Indian Buffet Game}, where each customer can request multiple dishes for a meal. It is worth pointing out that, there is an Indian Buffet process in machine learning, which defines a probability distribution over dishes for non-parametric learning methods \cite{ibp}. By introducing strategic behaviors into the non-strategic Indian Buffet process, the proposed Indian Buffet game is an ideal framework to study multiple dishes selection problem by incorporating social learning into strategic decision making with negative network externality. We will discuss two cases: Indian Buffet game without budget constraint and with budget constraint, where with budget constraint, the number of dishes each customer can require is limited, and vice versa. The main contributions of this paper can be summarized as follows.
\begin{enumerate}
\item We propose a general framework, Indian Buffet Game, to study how users make multiple concurrent selections under uncertain system state. Specifically, such a framework can reveal how users learn the uncertain system state through social learning and make optimal decisions to maximize their own expected utilities by considering negative network externality.

\item In the learning stage of the Indian Buffet Game, we propose a non-Bayesian social learning algorithm for customers to learn the dish states. Moreover, we prove theoretically the convergence of the proposed non-Bayesian social learning algorithm and show with simulations the fast convergence rate.

\item For the case without budget constraint, we show that the multiple concurrent dishes selection problem can be decoupled to a series of independent elementary Indian Buffet Game. We then design a recursive best response algorithm to find the subgame perfect Nash equilibrium of the elementary Indian Buffet Game. We show that, under the homogeneous setting, the Nash equilibrium profile exhibits a threshold structure.

\item For the case with budget constraint, we design a recursive best response algorithm to find the corresponding subgame perfect Nash equilibrium. We then show that, under the homogeneous setting, the Nash equilibrium profile exhibits an equal-sharing property.
\end{enumerate}

The rest of this paper is organized as follows. The system model is described in Section II. While, the Indian Buffet Game without budget constraint and with budget constraint are discussed in details in Section III and Section IV, respectively. In Section V, we give the theoretical proof of the convergence of the proposed non-Bayesian learning rule. Finally, we show simulation results in Section VI and draw conclusions in Section VII.

\section{System Model}

\subsection{Indian Buffet Game Formulation}
Let us consider an Indian buffet restaurant which provides $M$ dishes denoted by $r_1,r_2,...,r_M$. There are $N$ customers labeled with $1,2,...,N$ sequentially requesting dishes for meal. Each dish can be shared among multiple customers and each customer can select multiple dishes. We assume that all $N$ customers are rational in the sense that they will select dishes which can maximize their own utilities. In such a case, the multiple dishes selection problem can be formulated to be a non-cooperative game, called \textbf{Indian Buffet Game}, as follows:

\begin{itemize}
\item \emph{Players}: $N$ rational customers in the restaurant.
\item \emph{Strategies}: Since each customer can request multiple dishes, the strategy set can be defined as
    \begin{equation}
        \mathcal X=\Big\{\emptyset, \{r_1\},...,\{r_1,r_2\},...,\{r_1,r_2,...,r_M\}\Big\},
    \end{equation}
    where each strategy is a combination of dishes and $\emptyset$ means no dish is requested. Obviously, customers' strategy set is finite with $2^M$ elements. We denote the strategy of customer $i$ as $\mathbf d_i=\big(d_{i,1},d_{i,2},...,d_{i,M}\big)^\prime$\footnote{In the paper, the bold symbols represent vectors, the bold capital symbols represent matrixes, the subscript $i$ denotes the customer index, subscript $j$ denotes the dish index and the superscript $(t)$ denotes time slot index.}, where $d_{i,j} = 1$ represents customer $i$ requests dish $r_j$ and otherwise we have $d_{i,j} = 0$. The strategy profile of all customers can be denoted by a $M\times N$ matrix as follow
    \begin{equation}
        \!\!\!\mathbf D\!=\!(\mathbf d_1,\mathbf d_2,...,\mathbf d_N)\!=\!
        \begin{bmatrix}
            d_{1,1}&d_{2,1}&\cdots&d_{N,1}\\
            d_{1,2}&d_{2,2}&\cdots&d_{N,2}\\
            \vdots&\vdots&\ddots&\vdots\\
            d_{1,M}&d_{2,M}&\cdots&d_{N,M}
        \end{bmatrix}.\!\!\!\!\!\!\label{dmatrix}
    \end{equation}
\item \emph{Utility function}: The utility of each customer is determined by both the quality of the dish and the number of customers who share the same dish due to the negative network externality. The quality of one dish can be interpreted as the deliciousness or the size. Let $q_j\in Q$ denote the quality of dish $r_j$ where $Q$ is the quality space, and $N_j$ denote the total number of customers requesting dish $r_j$. Then, we can define the utility function of customer $i$ requesting dish $r_j$ as
    \begin{equation}
        u_{i,j}(q_j,N_j) = g_{i,j}(q_j,N_j) - c_{i,j}(q_j,N_j),\label{utility}
    \end{equation}
    where $g_{i,j}(\cdot)$ is the gain function and $c_{i,j}(\cdot)$ is the cost function. Note that the utility function is an increasing function in terms of $q_j$, and a decreasing function in terms of $N_j$, which can be regarded as the characteristic of negative network externality since the more customers request dish $r_j$, the less utility customer $i$ can obtain.
\end{itemize}

We here define the dish state $\bm \theta=\{\theta_1,\theta_2,...\theta_M\}$, where $\theta_j\in\Theta$ denotes the state of dish $r_j$. $\Theta$ is the set of all possible states, which is assumed to be finite. The dish state keeps unchanged along with time until the whole Indian buffet restaurant is remodeled. The aforementioned quality of dish $r_j$, $q_j$, is assumed to be a random variable following the distribution $f_j(\cdot|\theta_j)$, which means that the state of the dish $\theta_j$ determines the distribution of the dish quality $q_j$. The dish state $\bm \theta\in \Theta^M$ is unknown to all customers, i.e., they do not know exactly whether the dish is delicious or not before requesting. Nevertheless, they may have received some advertisements or gathered some reviews about the restaurant. Such information can be regarded as some kinds of signals related to the true state of the restaurant. In such a case, customers can estimate $\bm \theta$ through those available information, i.e., the information they know in advance and/or gather from other customers.

In the Indian Buffet Game model, we divide the system time into time slots and assume that the dish quality $q_j$ with $j=1,2,...,M$ varies independently from time slot to time slot following the corresponding conditional distributions $f_j(\cdot|\theta_j)$. In each time slot, customers make sequential decisions on which dishes to request. There are mainly two aspects needed to be addressed in the Indian Buffet Game. Firstly, since the states are unknown, it is very important to design an effective social learning rule for customers to learn from others and their previous outcomes. Secondly, given customers' knowledge about the states, we shall characterize the equilibrium that rational customers will adopt in each time slot. In this paper, to ensure fairness among customers, we assume that customers have different orders of selecting dishes at different time slots\footnote{In the sequel, the customer index $1,2,...,N$ means the dish request order of them, i.e., customer $i$ means the $i$-th customer.}. In such a case, it is sufficient for customers to only consider the expected utilities at current time slot.

Moreover, although each customer can request more than one dish, the total number of requests is subject to the following budget constraint:
\begin{equation}
\sum\limits_{j=1}^{M} d_{i,j} \le L, \ \ \ \forall\ i = 1,2,...,N.  \label{constraint}
\end{equation}
 A special case of (\ref{constraint}) is that $L\ge M$, which is equivalent to the case without budget constraint where customers can request as many dishes as possible. In Section III and IV, we will discuss the Indian Buffet Game under two scenarios: without budget constraint ($L\ge M$) and with budget constraint ($L< M$), respectively.

\subsection{Time Slot Structure of Indian Buffet Game}

Since the dish state $\bm \theta\in \Theta^M$ is unknown to customers, we introduce the concept of belief to describe customers' uncertainty about the state. Let us denote the belief as $\mathbf P^{(t)}=\big\{\mathbf p^{(t)}_j,j=1,2,..,M\big\}$, where $\mathbf p^{(t)}_j=\big\{p^{(t)}_j(\theta),\theta\in\Theta\big\}$ represents customers' estimation about the probability distribution of the state of dish $r_j$ at time slot $t$. Since customers can obtain some prior information about the dish state, we assume that all customers start with a prior belief $p^{(0)}_j(\theta)$ for every state $\theta_j$ . In this subsection, we will discuss the proposed social learning algorithm, i.e., how customers update their belief $\mathbf P^{(t)}$ at each time slot, and leave the convergence and performance analysis in Section V.

\begin{figure}[!t]
  \centering
  \centerline{\epsfig{figure=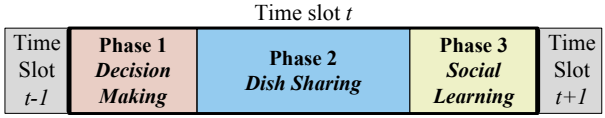,width=8cm}}
  \caption{Time slot structure of the Indian Buffet Game.}\label{slot}
\end{figure}

In Fig.\,\ref{slot}, we illustrate the time slot structure of the proposed Indian Buffet Game. At each time slot $t \in \{1,2...\}$, there are three phases: decision making phase, dish sharing phase and social learning phase.
\subsubsection{Phase 1 - decision making}
In this first phase, customers sequentially make decisions on which dishes to request and broadcast their decisions to others. For customer $i$, his/her decision is to maximize his/her expected utility at current time slot, based on the belief at current time slot $\mathbf P^{(t)}$, the decisions of the previous $(i-1)$ customers $\{\mathbf d_1,\mathbf d_2,...,\mathbf d_{i-1}\}$, and his/her predictions of the subsequent $(N-i)$ customers' decisions.
\subsubsection{Phase 2 - dish sharing} In the second phase, each customer requests his/her desired dishes and receives a utility $u_{i,j}(q_j,N_j)$ according to the dish quality $q_j$ and the number of customers $N_j$ sharing the same dish as defined in (\ref{utility}). Notice that since $N_j$ is known to all customers after the decision making phase, the customers who requested dish $r_j$ at time slot $t$ can infer the dish quality $q_j$ from the received utility. Let us denote such inferred information as $s_{i,j}^{(t)}\in Q, s_{i,j}\sim f_j(\cdot|\theta_j)$, which will serve as the signal in the learning procedure. On the other hand, the customers who have not requested $r_j$ at time slot $t$, cannot infer the dish quality $q_j$ and thus have no inferred signals. Such an asymmetric structure, i.e., not every customer receives signals, makes the learning problem different from the traditional social learning settings and thus poses more challenges on learning the true state.
\subsubsection{Phase 3 - social learning}
\begin{figure}[!t]
  \centering
  \centerline{\epsfig{figure=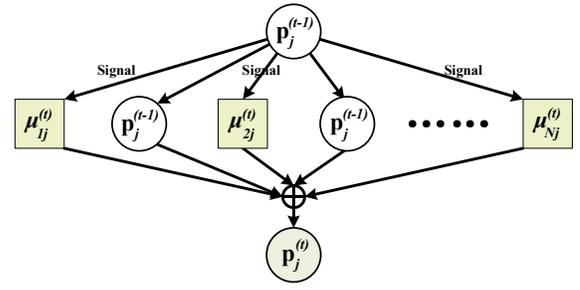,width=7.5cm}}
  \caption{Non-Bayesian learning rule for each dish.}\label{learning}
\end{figure}

With the observed/inferred signals in the second phase, customers can update belief through the proposed non-Bayesian social learning rule. As illustrated in Fig.\,\ref{learning}, there are mainly two steps in the proposed social learning rule. In the first step, each customer updates his/her local intermediate belief on $\theta_j$, $\bm \mu_{i,j}^{(t)}$, and then reveals this intermediate belief to others. In the second step, each customer combines his/her intermediate belief with others customers' intermediate beliefs in a linear manner\footnote{Note that the state learning processes, i.e., the belief update, of all dishes are independent.}. Based on the Bayes' theorem \cite{bayestheorem}, the customer $i$'s intermediate belief on the state of dish $r_j$, $\bm \mu_{i,j}^{(t)}=\{\mu_{i,j}^{(t)}(\theta),\theta\in\Theta\}$, can be calculated by
\begin{equation}
\mu_{i,j}^{(t)}(\theta)=\frac{f_j(s_{i,j}^{(t)}|\theta)p_j^{(t-1)}(\theta)}{\sum_\Theta {f_j(s_{i,j}^{(t)}|\theta)p_j^{(t-1)}(\theta)}}, \quad \forall\ \theta\in\Theta.\label{intermediate}
\end{equation}
From (\ref{intermediate}), we can see that when customer $i$ has requested $r_j$ at time slot $t$, he/she will incorporate the corresponding signal into his/her intermediate belief $\bm \mu_{i,j}^{(t)}$. Otherwise he/she will use the previous belief $\mathbf p_j^{(t)}$ instead. Then, each customer linearly combines his/her intermediate belief with others customers' intermediate beliefs as follow
\begin{align}
\label{NB1}
p_{j}^{(t)}(\theta) = &\frac{1}{N}\sum_{i=1}^N\left[d_{i,j}^{(t)}\mu_{i,j}^{(t)}(\theta) + \Big(1-d_{i,j}^{(t)}\Big)p_j^{(t-1)}(\theta)\right], \\
& \forall\ \theta\in\Theta, \text{ and } j =1,2,...,M, \nonumber
\end{align}
where $d_{i,j}^{(t)}$ is the strategy of customer $i$ at time slot $t$.

\section{Indian Buffet Game without Budget Constraint}

In this section, we study the Indian Buffet Game without budget constraint, which is corresponding to the case where $L\ge M$ in (\ref{constraint}). When there is no budget constraint, customers should request all dishes that can give them positive expected utility to maximize their total expected utilities. We will first show that without budget constraint, whether to request a dish is independent of other dishes, i.e. the Indian Buffet Game that select multiple concurrent dishes is decoupled to a series of elementary Indian Buffet Game that select a single dish. Then we present a recursive algorithm that characterize the subgame perfect equilibrium of the Indian Buffet Game without budge constraint. Finally, we discuss the homogeneous case where customers have the same form of utility function to gain more insights.

To show the independence among different dishes, we first define the best response of a customer given other customers' actions. Let us define $\mathbf{n}_{-i} = \{n_{-i,1},n_{-i,2},...,n_{-i,M}\}$ with
\begin{equation}
n_{-i,j} = \sum\limits_{k \ne i} d_{k,j}
\end{equation}
being the number of customers except customer $i$ choosing $r_j$. Let $\mathbf P = \{\mathbf p_1,\mathbf p_2,...,\mathbf p_M\}$, where $\mathbf p_j=\{p_j(\theta),\theta\in\Theta\}$ is customers' belief regarding the state of dish $r_j$ at current time slot\footnote{Since we discuss the Indian Buffet Game in one time slot, the superscript $(t)$ is omitted in Section III and IV.}. Given $\mathbf P$ and $\mathbf{n}_{-i}$, the best response of customer $i$, $\mathbf{d}_i^* = (d_{i,1}^*, d_{i,2}^*,...,d_{i,M}^*)^\prime$, can be written as
\begin{equation}
\label{bestR1}
\mathbf{d}_i^* = \mbox{BR}_i(\mathbf P,\mathbf{n}_{-i})=\mathop {\arg \max }\limits_{\mathbf{d}_i \in \{ 0,1\}^M } \sum\limits_{j = 1}^M {d_{i,j}}\cdot U_{i,j},
\end{equation}
where $U_{i,j}$ is customer $i$'s expected utility of requesting dish $r_j$ given belief $\mathbf P$, which can be calculated by
\begin{equation}
\label{exutility}
U_{i,j}=\sum_\Theta  {\sum_Q {u_{i,j}({q_j},{n_{ - i,j}} + {d_{i,j}})f_j ({q_j}|{\theta _j}){p_j}({\theta _j})} },
\end{equation}
where $Q$ is the quality/signal set and $q_j\in Q$.

From (\ref{bestR1}) and (\ref{exutility}), we can see that the optimal decision on one dish is irrelevant to the decisions on others, which leads to the independence among different dishes. In such a case, we have
\begin{equation}
d_{i,j}^* = \mathop {\arg \max }\limits_{d_{i,j} \in \{ 0,1\} } d_{i,j}\cdot U_{i,j}.
\end{equation}
The independence property enables us to simplify our analysis by breaking the origin Indian Buffet Game into $M$ elementary Indian Buffet Game, each of which involves only one dish selection. In the remaining of this section, we will focus on the analysis of the elementary Indian Buffet Game and drop the dish index $j$ for notation simplification. As a result, we can rewrite the best response of customer $i$ as
\begin{align}
d_{i}^* &= \mbox{BR}_i(\mathbf p,n_{-i}) = \mathop {\arg \max }\limits_{d_{i} \in \{ 0,1\} } d_i \cdot U_i \label{bestR2}\\
&=\left\{ {\begin{array}{*{20}{c}}
1,  &\text{  if  } {U_i\!=\!\sum\limits_\Theta  {\sum\limits_Q {u_{i}({q},{n_{ - i}} + 1)f ({q}|{\theta}){p}({\theta})} } } \! > \!0;\\
0,  &\text{  otherwise.  }
\end{array}} \right.  \nonumber
\end{align}

\subsection{Recursive Best Response Algorithm}

In this subsection, we study how to solve the best response defined in (\ref{bestR2}) for each customer. From (\ref{bestR2}), we can see that customer $i$ needs to know $n_{-i}$ to calculate the expected utility $U_i$ to decide whether to request the dish or not. However, since customers make decisions sequentially, customer $i$ does not know the decisions of those who are after him/her and thus needs to predict the subsequent customers' decisions based on the belief and known information.

Let $m_i$ denote the number of customers that will request the dish after customer $i$, then we can write the recursive form of $m_i$ as
\begin{equation}
m_i=m_{i+1}+d_{i+1}.
\end{equation}
Let $m_i|_{d_i = 0}$ and $m_i|_{d_i = 1}$ represent $m_i$ under the condition of $d_i = 0$ and $d_i = 1$, respectively. Denote by $n_i = \sum\limits_{k=1}^{i-1} d_k$, the number of customers choosing the dish before customer $i$. Then, the estimated number of customers choosing the dish excluding customer $i$ can be written as follows:
\begin{align}
\hat{n}_{-i}|_{d_i = 0} &= n_i + m_i|_{d_i = 0},  \\
\hat{n}_{-i}|_{d_i = 1} &= n_i + m_i|_{d_i = 1}.\label{n-i}
\end{align}
Note that $\hat{n}_{-i}|_{d_i = 0}$ and $\hat{n}_{-i}|_{d_i = 1}$ are different from $n_{-i}$ in that the values of $d_{i+1},d_{i+2},...,d_{N}$ are estimated instead of true observations.

According to (\ref{n-i}), we can compute the expected utility of customer $i$ when $d_i=1$ as
\begin{equation}
U_{i}|_{d_i=1} = {\sum_\Theta  {\sum_Q {u_{i}({q},{n_{i}} + m_i|_{d_i = 1} + 1)f ({q}|{\theta}){p}({\theta})} } } .
\end{equation}
Since the utility of customer $i$ is zero when $d_i=0$, the best response of customer $i$ can be obtained as
\begin{equation}
\label{bestR3}
d_i^* = \left\{ {\begin{array}{*{20}{c}}
1,  &\text{  if  }\ U_{i}|_{d_i=1} > 0;\\
0,  &\text{  otherwise.  }
\end{array}} \right.
\end{equation}

With (\ref{bestR3}), we can find the best response of customer $i$ given belief $\mathbf p$, current observation $n_i$ and predicted number of subsequent customers choosing the dish, $m_i|_{d_i = 1}$. To predict $m_i|_{d_i = 1}$, customer $i$ needs to predict the decisions of all customers from $i+1$ to $N$. When it comes to customer $N$, since he/she knows exactly the decisions of all the previous customers, he/she can find the best response without making any prediction, i.e., $m_N=0$. Along this line, it is intuitive to design a recursive algorithm to predict $m_i|_{d_i = 1}$ by considering all possible decisions of customers from $i+1$ to $N$ and updating $m_i=m_{i+1}+d_{i+1}$. In Algorithm 1, we show the recursive algorithm $\mbox{BR\_EIBG}(\mathbf{p},n_i,i)$ that describes how to predict $m_i|_{d_i = 1}$ and find the best response $d_i$ for customer $i$, given current belief $\mathbf p$ and observation $n_i$. Moreover, in order to give a correct prediction of $m_i$ in the recursion procedure, we calculate and return $m_i|_{d_i = 0}$ when the best response of customer $i$ is $0$. In the following, we will prove that the action profile specified in $\mbox{BR\_EIBG}(\mathbf{p},n_i,i)$ is a subgame perfect Nash equilibrium for the elementary Indian Buffet Game.

\begin{algorithm}
\caption{$\mbox{\mbox{BR\_EIBG}}(\mathbf{p},n_i,i)$} \label{alg2}
\begin{algorithmic}
\IF {Customer $i == N$}
    \STATE //******\textbf{For customer $N$}******//
    \IF {$U_N={\sum\limits_\Theta  {\sum\limits_Q {u_{N}({q},{n_{N}} + 1)f ({q}|{\theta})p({\theta})} } }  > 0$}
        \STATE $d_N \gets 1$
    \ELSE
        \STATE $d_N \gets 0$
    \ENDIF
    \STATE $m_N \gets 0$
\ELSE
    \STATE //******\textbf{For customer $1,2,...,N-1$}******//
    \STATE //***\emph{Predicting}***//
    \STATE $(d_{i+1},m_{i+1}) \gets \mbox{BR\_EIBG}(\mathbf{p},n_i + 1, i+1)$
    \STATE $m_i \gets m_{i+1} + d_{i+1}$
    \STATE //***\emph{Making decision}***//
    \IF {${U_i=\sum\limits_\Theta  {\sum\limits_Q {u_{i}({q},{n_{i}} + m_i + 1)f ({q}|{\theta})p({\theta})} } }  > 0$}
        \STATE $d_i \gets 1$
    \ELSE
        \STATE $(d_{i+1},m_{i+1}) \gets \mbox{BR\_EIBG}(\mathbf{p},n_i, i+1)$
        \STATE $m_i \gets m_{i+1} + d_{i+1}$
        \STATE $d_i \gets 0$
    \ENDIF
\ENDIF
\RETURN $(d_i, m_i)$

\end{algorithmic}
\end{algorithm}
\subsection{Subgame Perfect Nash Equilibrium}

In this subsection, we will show that Algorithm \ref{alg2} leads to the subgame perfect Nash equilibrium for the elementary Indian Buffet Game. In the following, we first give the formal definitions of Nash equilibrium, subgame and subgame perfect Nash equilibrium as follows.
\begin{definition}
Given the belief $\mathbf{p}=\{p(\theta),\theta\in\Theta\}$, the action profile $\mathbf{d}^* = \{d_1^*,d_2^*,...,d_N^*\}$ is a Nash equilibrium of the $N$-customer elementary Indian Buffet Game if and only if $\forall\ i\in\{1,2,...,N\}$, $d_i^* = \mbox{BR}_i\bigg(\mathbf p,\sum\limits_{k \ne i} d_k^*\bigg)$ as given in (\ref{bestR2}).
\end{definition}

\begin{definition}
A subgame of the $N$-customer elementary Indian Buffet Game consists of the following three elements: 1) it starts from customer $i$ with $i=1,2,...,N$; 2) it has the belief at current time slot, $\mathbf p$; 3) it has current observation, $n_i$, which are the decisions of previous customers.
\end{definition}

\begin{definition}
A Nash equilibrium is a subgame perfect Nash equilibrium if and only if it is a Nash equilibrium for every subgame.
\end{definition}

With the above definitions, we show in the following theorem that the action profile derived by Algorithm \ref{alg2} is a subgame perfect Nash equilibrium of the elementary Indian Buffet Game.
\begin{theorem}
Given the belief $\mathbf{p}=\{p(\theta),\theta\in\Theta\}$, the action profile $\mathbf{d}^* = \{d^*_1,d^*_2,...,d^*_N\}$, with $d^*_i$ being determined by $\mbox{BR\_EIBG}(\mathbf{p},n_i,i)$ and $n_i = \sum\limits_{k=1}^{i-1} d^*_k$, is a subgame perfect Nash equilibrium for the elementary Indian Buffet Game.
\end{theorem}
\begin{IEEEproof}
We first show that $d^*_k$ is the best response of customer $k$ in the subgame starting from customer $i, \forall\ 1\leq i\leq k\leq N$.

If $k=N$, we can see that $\mbox{BR\_EIBG}(\mathbf{p},n_N,N)$ assigns the value of $d^*_N$ directly as
\begin{equation}
d^*_N = \left\{ {\begin{array}{*{20}{c}}
1, &\!\!\!\!\text{  if  } {U_N\!=\!\sum\limits_\Theta  {\sum\limits_Q {u_{N}({q},{n_{N}} + 1)f ({q}|{\theta})p({\theta})} } } \! > \!0;\\
0, &\text{  otherwise.  }
\end{array}} \right.\!\!\!\!\!\!\!
\end{equation}
Since $n_{N} = n_{-N}$, we have $d^*_k = \mbox{BR}_k(\mathbf p,n_{-k})$ in the case of $k=N$ according to (\ref{bestR2}), i.e. $d^*_k$ is the best response of customer $k$.

If $k<N$, suppose $d^*_k$ is the best response of customer $k$ derived by $\mbox{BR\_EIBG}(\mathbf{p},n_k,k)$. If $d^*_k=0$, denoting $d^\prime_k=1$ as the contradiction, we can see from $\mbox{BR\_EIBG}(\mathbf{p},n_k,k)$ that
\begin{equation}
U_k|_{d^*_k=1}\!\!=\!\!\sum\limits_\Theta  {\sum\limits_Q {u_{k}({q},{n_{k}}\! +\! m_k\! +\! 1)f ({q}|{\theta})p({\theta})} }   > 0=U_k|_{d^\prime_k=0},
\end{equation}
which means that customer $k$ has no incentive to deviate from $d^*_k=1$ given the prediction of other customers' decisions. If $d^*_k=1$, denoting $d^\prime_k=0$ as the contradiction, we can see from $\mbox{BR\_EIBG}(\mathbf{p},n_k,k)$ that
\begin{equation}
U_k|_{d^\prime_k=0}=0>U_k|_{d^*_k=1}\!\!=\!\!\sum\limits_\Theta  {\sum\limits_Q {u_{k}({q},{n_{k}}\! +\! m_k\! +\! 1)f ({q}|{\theta})p({\theta})} },
\end{equation}
which means that customer $k$ has no incentive to deviate from $d^*_k=0$ given the prediction of other customers' decisions. Therefore, $d^*_k = \mbox{BR\_EIBG}(\mathbf{p},n_k,k)$ is the best response of customer $k$ in the subgame of the elementary Indian Buffet Game starting with customer $i$. Moreover, since the statement is true for $\forall\  k$ satisfying $i\le k\le N$, we know that $\{d^*_i,d^*_{i+1},...,d^*_{N}\}$ is the Nash equilibrium for the subgame starting from customer $i$. Finally, according to the definition of subgame perfect Nash equilibrium, we can conclude that \emph{Theorem 1} is true.
\end{IEEEproof}

\subsection{Homogeneous Case}
From the previous subsection, we know that a recursive procedure is needed to determine the best responses of the elementary Indian Buffet Game. This is due to the fact that we need to predict the decisions of all subsequent customers to determine the best response of a certain customer. In this subsection, we simplify the game with homogeneous setting to derive more concise best response.

In the homogeneous case, we assume that all customers have the same form of utility function, i.e., $u_i(q,n)=u(q,n)$, for all $i,q,n$. Under such a setting, the equilibrium can be characterized in a much simpler way.

\begin{lemma}
In the $N$-customer elementary Indian Buffet Game under homogeneous settings, if $\mathbf{d^*} = \{d^*_1,d^*_2,...,d^*_N\}$ is the Nash equilibrium action profile specified by $\mbox{BR\_EIBG}()$, then we have $d^*_i = 1$ if and only if $0\le i\le n^*$, where $n^* = \sum\limits_{k=1}^N d^*_k$.
\end{lemma}
\begin{IEEEproof}
Suppose the best response of customer $i$, $d^*_i = 0$. Then, according to Algorithm 1, we have
\begin{equation}
U_i={\sum_\Theta  {\sum_Q {u({q},{n_{i}} + m_i|_{d_i = 1} + 1)f ({q}|{\theta}){p}({\theta})} } }  \le 0.\label{17}
\end{equation}
The prediction of $m_i$ under the condition of $d_i = 1$ relies on the recursive estimations of all subsequent customers' decisions. In particular, we have $m_i|_{d_i=1} = d_{i+1}|_{d_i=1} + m_{i+1}|_{d_i=1}$, where the value of $d_{i+1}|_{d_i=1}$ can be computed as follows
\begin{equation}
d_{i+1}|_{d_i=1} = \left\{ {\begin{array}{*{20}{c}}
1,  &\text{  if  }\ U_{i+1}|_{d_i=1}> 0;\\
0,  &\text{  otherwise,  }
\end{array}} \right.
\end{equation}
with
\begin{equation}
U_{i+1}|_{d_i=1}=\sum_\Theta  {\sum_Q {u({q},{n_{i}} + 1 + m_{i+1}|_{d_i = 1} + 1)f ({q}|{\theta}){p}({\theta})} }.\label{19}
\end{equation}
Since $n_{i} + 1 + m_{i+1}|_{d_i = 1} + 1 \ge n_{i} + m_i|_{d_i = 1} + 1$ and $u(q,n)$ is a decreasing function in terms of $n$, we have $d_{i+1}|_{d_i=1} = 0$ according to (\ref{17}) and (\ref{19}). Following the same argument, we can show that $d_{k}|_{d_i=1} = 0$ for all $k = i+1,i+2,...,N$. Therefore, we have
\begin{equation}
m_i|_{d_i = 1} = \sum\limits_{k=i+1}^N d_k|_{d_i = 1} = 0.
\end{equation}

Then, let us consider the best response of customer $i+1$, which can be calculated by
\begin{equation}
d^*_{i+1} = \left\{ {\begin{array}{*{20}{c}}
1,  &\text{  if  }\ U_{i+1} > 0;\\
0,  &\text{  otherwise.  }
\end{array}} \right.
\end{equation}
where
\begin{equation}
U_{i+1}=\sum_\Theta  {\sum_Q {u({q},{n_{i+1}} + m_{i+1}|_{d_{i+1} = 1} + 1)f ({q}|{\theta}){p}({\theta})} } .\label{22}
\end{equation}
Since $n_{i+1} = n_{i} + d_i$, $m_i|_{d_i=1}=0$ and $m_{i+1}|_{d_{i+1} = 1} \ge 0$, we have $n_{i+1} + m_{i+1}|_{d_{i+1} = 1} + 1 \ge {n_{i}} + m_i|_{d_i = 1} + 1$. According to (\ref{17}), (\ref{22}) and the decreasing property of utility function in terms of number of customers sharing the same dish, we have $d^*_{i+1} = 0$.

Following the same argument, we can show that if $d^*_i = 0$, then $d^*_k = 0$ for all $k\in\{ i+1,i+2,...,N\}$. Since all decisions can take values of either $0$ or $1$, we have $d^*_i=1$ if and only if $0\le i\le \sum\limits_{k=1}^N d^*_k$. This completes the proof.
\end{IEEEproof}

From \emph{Lemma 1}, we can see that there exists a threshold structure in the Nash equilibrium of elementary India Buffet Game with homogeneous setting. The threshold structure is embodied in the fact that if $d^*_i=0$, then $d^*_k=0,\forall\ k\in\{i+1,i+2,...,N\}$, and if $d^*_i=1$, then $d^*_k=1,\forall\ k\in\{1,2,...i-1\}$. The result can be easily extended to the Indian Buffet Game without budget constraint under the homogeneous setting as shown in the following theorem.
\begin{theorem}
In the $M$-dish and $N$-customer Indian Buffet Game without budget constraint, if all the customers have same utility functions, there exists a threshold structure in the Nash equilibrium matrix $\mathbf D^*$ denoted by (\ref{dmatrix}), i.e., for any row $j\in\{1,2,...,M\}$ of $\mathbf D^*$, there is a $T_j\in\{1,2,...,N\}$ satisfying that
\begin{equation}
d^*_{i,j}=\left\{\begin{array}{ll}
1,&\forall \ i< T_j; \\
0,&\forall\ i\ge T_j.
\end{array}\right.
\end{equation}

\end{theorem}
\begin{IEEEproof}
This theorem directly follows by extending \emph{Lemma 1} into $M$ independent dishes case.
\end{IEEEproof}

\section{Indian Buffet Game with Budget Constraint}

In this section, we study the Indian Buffet Game with budget constraint, which is corresponding to the case with $L< M$ in (\ref{constraint}). Unlike previous case, when there is budget constraint for each customer, the selection among different dishes is no longer independent but coupled. In the following, we will first discuss a recursive algorithm that can characterize the subgame perfect Nash equilibrium of the Indian Buffet Game with budget constraint. Then, we discuss a simplified case with homogeneous setting to gain more insights.

\subsection{Recursive Best Response Algorithm}

In the budget constraint case, we assume that each customer can at most request $L$ dishes at each time slot with $L<M$. In such a case, the best response of customer $i$ can be found by the following optimization problem.
\begin{eqnarray}
\label{bestR11}
&\mathbf{d}_i^* = \mbox{BR}_i(\mathbf P,\mathbf{n}_{-i})=\mathop {\arg \max }\limits_{\mathbf{d}_i \in \{ 0,1\}^M } \sum\limits_{j = 1}^M {d_{i,j}}\cdot U_{i,j},&\\
&\mbox{s.t.}\ \sum\limits_{i=1}^{N}d_{i,j}\le L<M,&\nonumber
\end{eqnarray}
where
\begin{equation}
U_{i,j}=\sum_\Theta  {\sum_Q {u_{i,j}({q_j},{n_{ - i,j}} + {d_{i,j}})f_j ({q_j}|{\theta _j}){p_j}({\theta _j})} }.
\end{equation}
From (\ref{bestR11}), we can see that customer $i$'s decision on dish $r_j$ is coupled with all other dishes, and thus (\ref{bestR11}) cannot be decomposed into $M$ subproblems. Nevertheless, we can still find the best response of each customer by comparing all possible combinations of $L$ dishes. Let $\mathbf \Phi=\{\bm \phi_1,\bm \phi_2,...,\bm \phi_H\}$ denote the set of all combinations of $l\ (1\le l\le L)$ dishes out of $M$ dishes, where $H=\sum\limits_{l=1}^LC_M^l=\sum\limits_{l=1}^L\frac{M!}{l!(M-l)!}$ and $\bm \phi_h =(\phi_{h,1},\phi_{h,2},...,\phi_{h,M})^\prime$ is one possible combination with $\phi_{h,j}$ representing whether dish $r_j$ is requested, e.g.,
\begin{equation}
\bm \phi_h=(\underbrace{1,1,...,1}_{l},\underbrace{0,0,...,0}_{M-l})^\prime
\end{equation}
means the customer requests dish $r_1,r_2,...,r_l\ (1\le l\le L)$. In other words, $\mathbf \Phi$ is the candidate strategy set of each customer with constraint $L$.

Let us define customer $i$'s observation of previous customers' decisions as
\begin{equation}
\mathbf n_i=\{n_{i,1},n_{i,2},...,n_{i,M}\},
\end{equation}
where $n_{i,j}=\sum\limits_{k=1}^{i-1}d_{k,j}$ is the number of customers choosing dish $r_j$ before customer $i$. Let $\mathbf m_i$ denote the subsequent customers' decisions after customer $i$, we have its recursive form as
\begin{equation}
\mathbf m_i=\mathbf m_{i+1}+\mathbf d_{i+1}.
\end{equation}
Then, let
\begin{equation}
\mathbf m_i|_{\mathbf d_i=\bm \phi_h}=\{m_{i,1}|_{\mathbf d_i=\bm \phi_h},m_{i,2}|_{\mathbf d_i=\bm \phi_h},...,m_{i,M}|_{\mathbf d_i=\bm \phi_h}\},
\end{equation}
with $m_{i,j}|_{\mathbf d_i=\bm \phi_h}$ being the predicted number of subsequent customers that will request dish $r_j$ under the condition that $\mathbf d_i=\bm \phi_h$, where $\mathbf d_i=(d_{i,1},d_{i,1},...,d_{i,M})^\prime$ and $\bm \phi_h\in\mathbf\Phi$. In such a case, the predicted number of customers choosing different dishes excluding customer $i$ is
\begin{equation}
\hat{\mathbf n}_{-i}|_{\mathbf d_i = \bm \phi_h} = \mathbf n_{i} + \mathbf m_{i}|_{\mathbf d_i = \bm \phi_h}.
\end{equation}

According to above definitions, we can write customer $i$'s expected utility by obtaining dish $r_j$ when $\mathbf d_i=\bm \phi_h$ as
\begin{eqnarray}
U_{i,j}|_{\mathbf d_i=\bm \phi_h}\!\!\!\!& = &\!\!\!\!\phi_{h,j} \sum_\Theta  \sum_Q u_{i,j}(q_j, n_{i,j}+m_{i,j}|_{\mathbf d_i = \bm \phi_h}+ \phi_{h,j}) \nonumber\\
&&\!\!\!\!f_j({q_j}|{\theta_j}){p_j}({\theta_j}).\label{uij}
\end{eqnarray}
Then, the total expected utility customer $i$ can obtain with $\mathbf d_i=\bm \phi_h$ is the sum of $U_{i,j}|_{\mathbf d_i=\bm \phi_h}$ over all $M$ dishes, i.e.,
\begin{equation}
U_i|_{\mathbf d_i=\bm \phi_h}=\sum\limits_{j=1}^M U_{i,j}|_{\mathbf d_i=\bm \phi_h}.
\end{equation}
In such a case, we can find the optimal $\bm\phi^*_h$ which maximizes customer $i$'s expected utility $U_i|_{\mathbf d_i=\bm \phi_h}$ as follow
\begin{equation}
\bm \phi^*_h=\mathop{\arg\max}_{\bm \phi_h\in\mathbf \Phi} \{U_{i}|_{\mathbf d_i=\bm \phi_h}\},\label{bestR4}
\end{equation}
which is the best response of customer $i$.

To obtain the best response in (\ref{bestR4}), each customer needs to calculate the expected utilities defined in (\ref{uij}), which in turn requires to predict $m_{i,j}|_{\mathbf d_i=\bm \phi_h}$, i.e., the number of customers who choose dish $r_j$ after customer $i$. When it comes to customer $N$ who has already known all previous customers' decisions, no prediction is required. Therefore, similar to Algorithm 1, given belief $\mathbf{P} = \{\mathbf p_1,\mathbf p_2,...,\mathbf p_M\}$ at current time slot and current observation $\mathbf n_i=\{n_{i,1},n_{i,2},...,n_{i,M}\}$, we design another recursive best response algorithm $\mbox{BR\_IBG}(\mathbf p,\mathbf n_i,i)$ for solving the Indian Buffet Game with budget constraint in Algorithm 2. As we can see, customer $N$ only needs to compare the expected utilities of requesting all $M$ dishes and choose $L$ or less than $L$ dishes with highest positive expected utilities. Note that $\max^L$ means finding the highest $L$ values. For other customers, each one needs to first recursively predict the following customers' decisions, and then make his/her own decision based on the prediction and current observations.

\begin{algorithm}
\caption{$\mbox{\mbox{BR\_IBG}}(\mathbf{P},\mathbf n_i,i)$} \label{alg3}
\begin{algorithmic}
\IF {Customer $i == N$}
    \STATE //******\textbf{For customer $N$}******//
    \FOR {$j=1$ to $M$}
        \STATE $U_{i,j}={\sum\limits_\Theta  {\sum\limits_Q {u_{N,j}({q_j},{n_{N,j}} + 1)f_j ({q_j}|{\theta_j})p_j({\theta_j})} } }$
    \ENDFOR
    \STATE $\mathbf j=\{j_1,j_2,...,j_L\}\gets \mathop {\arg \max^L}\limits_{j\in\{1,2,...,M\}}\{U_{i,j}\}$
    \FOR {$j=1$ to $M$}
        \IF {$(U_{i,j}>0)$\&\&$(j\in\mathbf j)$}
            \STATE $d_{N,j} \gets 1$
        \ELSE
            \STATE $d_{N,j} \gets 0$
        \ENDIF
    \ENDFOR
    \STATE $\mathbf m_N=\mathbf 0$
\ELSE
    \STATE //******\textbf{For customer $1,2,...,N-1$}******//
    \STATE //***\emph{Predicting}***//
    \FOR {$\bm\phi_h=\bm\phi_1$ to $\bm\phi_H$}
        \STATE $(\mathbf d_{i+1},\mathbf m_{i+1}) \gets \mbox{BR\_IBG}(\mathbf{P},\mathbf n_i + \bm\phi_h, i+1)$
        \STATE $\mathbf m_i \gets \mathbf m_{i+1} + \mathbf d_{i+1}$
        \STATE $U_{i}(\phi_h)=\sum\limits_M\phi_{h,j}\sum\limits_\Theta \sum\limits_Q u_{i,j}({q_j},{n_{i,j}} + m_{i,j} + \phi_{h,j})$
        \STATE \quad\quad\quad\quad$\cdot f_j ({q_j}|{\theta_j})p_j({\theta_j})$
    \ENDFOR
    \STATE //***\emph{Making decision}***//
    \STATE $\bm \phi^*_h\gets \mathop {\arg \max}\limits_{\bm \phi_h\in\mathbf\Phi}\{U_{i}(\bm\phi_h)\}$
    \STATE $(\mathbf d_{i+1},\mathbf m_{i+1}) \gets \mbox{BR\_IBG}(\mathbf{P},\mathbf n_i + \bm \phi^*_h, i+1)$
    \STATE $\mathbf d_i\gets \bm \phi^*_h$
    \STATE $\mathbf m_i \gets \mathbf m_{i+1} + \mathbf d_{i+1}$
\ENDIF
\RETURN $(\mathbf d_i, \mathbf m_i)$
\end{algorithmic}
\end{algorithm}
\subsection{Subgame Perfect Nash Equilibrium}

Similar to the elementary Indian Buffet Game, we first give formal definitions of the Nash equilibrium and subgame of Indian Buffet Game with budget constraint.
\begin{definition}
Given the belief $\mathbf{P}=\{\mathbf p_1,\mathbf p_2,...,\mathbf p_M\}$, the action profile $\mathbf{D}^* = \{\mathbf d_1^*,\mathbf d_2^*,...,\mathbf d_N^*\}$ is a Nash equilibrium of the $M$-dish and $N$-customer Indian Buffet Game with budget constraint $L$, if and only if $\mathbf d_i^* = \mbox{BR}_i\bigg(\mathbf P,\sum\limits_{k \ne i} \mathbf d_k^*\bigg)$ as defined in (\ref{bestR11}) for all $i$.
\end{definition}

\begin{definition}
A subgame of the $M$-dish and $N$-customer Indian Buffet Game with budget constraint $L$ consists of the following three elements: 1) it starts from customer $i$ with $i=1,2,...,N$; 2) it has the belief at current time slot, $\mathbf P$; 3) it has current observation, $\mathbf n_i$, which are the decisions of previous customers.
\end{definition}

Based on \emph{Definition 3, 4 and 5}, we show in the following theorem that the action profile obtained by Algorithm 2 is a subgame perfect Nash equilibrium of the Indian Buffet Game with budget constraint.
\begin{theorem}
Given the belief $\mathbf{P}=\{\mathbf p_1,\mathbf p_2,...,\mathbf p_M\}$, the action profile $\mathbf{D}^* = \{\mathbf d^*_1,\mathbf d^*_2,...,\mathbf d^*_N\}$, where $\mathbf d^*_i$ determined by $\mbox{BR\_IBG}(\mathbf{P},\mathbf n_i,i)$ and $\mathbf n_i = \sum\limits_{k=1}^{i-1} \mathbf d^*_k$, is a subgame perfect Nash equilibrium for the elementary Indian Buffet Game.
\end{theorem}
\begin{IEEEproof}
The proof of this theorem is similar to that of \emph{Theorem 1}, the details of which are omitted due to page limitation. The proof outline is that first to show $\forall\ i,k$ such that $ 1 \le i \le N$ and $i \le k \le N$, $\mathbf d^*_k$ is the best response of customer $k$ in the subgame starting from customer $i$ by analyzing two cases: $k=N$ and $k<N$. Then, we can know that $\{\mathbf d^*_i,\mathbf d^*_{i+1},...,\mathbf d^*_{N}\}$ is the Nash equilibrium for the subgame starting from customer $i$. Finally, according to the definition of subgame perfect Nash equilibrium, we can conclude that \emph{Theorem 3} is true.
\end{IEEEproof}

\subsection{Homogenous Case}

In the homogenous case, we assume that all customers' utility functions are the same, i.e., $u_{i,j}(q,n)=u(q,n)$; and all dishes are in the same state, i.e., the dish state $\bm \theta=\{\theta,\theta,...,\theta\}$. Under such circumstances, we can find some special property in the Nash equilibrium action profile $\mathbf D^*$ of the Indian Buffet Game with budget constraint. First, let us define a parameter $n_T$ which satisfies
\begin{equation}
\left\{\begin{array}{cc}\sum\limits_\Theta \sum\limits_Q u(q,n)f(q|\theta)p(\theta)>0,&\mbox{if } n\le n_T;\\
\sum\limits_\Theta \sum\limits_Q u(q,n)f(q|\theta)p(\theta)\le0,&\mbox{if } n> n_T.\end{array}\right.\label{nthreshold}
\end{equation}
From (\ref{nthreshold}), we can see that $n_T$ is the critical value such that the utility of $n_T$ customers sharing a certain dish is positive but becomes non-positive with one extra customer, i.e., each dish can be requested by at most $n_T$ customers. In the following theorem, we will show that, under the homogeneous setting, all dishes will be requested by nearly equal number of customers, i.e., the equal-sharing is achieved.
\begin{theorem}
In the  $M$-dish and $N$-customer Indian Buffet Game with budget constraint $L$, if all $M$ dishes are in the same states and all $N$ customers have the same utility function, the Nash equilibrium matrix $\mathbf D^*$ denoted by (\ref{dmatrix}) satisfies that, for all dishes $\{r_j, j=1,2,...,M\}$,
\begin{equation}
\sum\limits_{i=1}^Nd^*_{i,j}=\left\{\begin{array}{cc}
n_T,&\quad \mbox{if }\ n_T\le \left\lfloor\frac{NL}{M}\right\rfloor;\\&\\
\left\lfloor\frac{NL}{M}\right\rfloor\mbox{ or }\left\lceil\frac{NL}{M}\right\rceil,&\quad \mbox{if }\ n_T\ge \left\lceil\frac{NL}{M}\right\rceil.
\end{array}\right.\label{homo2}
\end{equation}
\end{theorem}

\begin{IEEEproof}
We prove this theorem by contradiction as follows.

\begin{itemize}
\item Case 1: $n_T\le \left\lfloor\frac{NL}{M}\right\rfloor$.
\end{itemize}

Suppose that there exists a Nash equilibrium $\mathbf D^{*}$ that contradicts with (\ref{homo2}). That is, there is a dish $r_{j^\prime}$ such that $\sum\limits_{i=1}^Nd^*_{i,j^\prime}> n_T$ or $\sum\limits_{i=1}^Nd^*_{i,j^\prime}< n_T$. From (\ref{nthreshold}), we know that each dish can be requested by at most $n_T$ customers, which means that only $\sum\limits_{i=1}^Nd^*_{i,j^\prime}< n_T$ may hold. If $\sum\limits_{i=1}^Nd^*_{i,j^\prime}< n_T\le \left\lfloor\frac{NL}{M}\right\rfloor$, we have $\sum\limits_{j=1}^M\sum\limits_{i=1}^Nd^*_{i,j}<NL$, which means that there exists at least one customer $i^\prime$ that requests less than $L$ dishes, i.e., $\sum\limits_{j=1}^M d^*_{i^\prime j}<L$. However, according to (\ref{nthreshold}), we have $\sum\limits_\Theta \sum\limits_Q u\Big(q,\sum\limits_{i=1}^Nd^*_{i,j^\prime}+1\Big)f(q|\theta)p(\theta)>0$, which means that the utility of customer $i^\prime$ can increase if he/she requests dish $r_{j^\prime}$, i.e., his/her utility is not maximized unless $\mathbf D^{*}$ is not a Nash equilibrium. This contradicts with our assumption. Therefore, we have $\sum\limits_{i=1}^Nd^*_{i,j}= n_T$ for all dishes when $n_T\le \left\lfloor\frac{NL}{M}\right\rfloor$.

\begin{itemize}
\item Case 2: $n_T\ge \left\lceil\frac{NL}{M}\right\rceil$.
\end{itemize}

Similar to the arguments in case 1, we cannot have $\sum\limits_{j=1}^M\sum\limits_{i=1}^Nd^*_{i,j}<NL$, which means that $\sum\limits_{j=1}^M\sum\limits_{i=1}^Nd^*_{i,j}=NL$. Let us assume that there exists a Nash equilibrium $\mathbf D^{*}$ that contradicts with (\ref{homo2}). Since $\sum\limits_{j=1}^M\sum\limits_{i=1}^Nd^*_{i,j}=NL$, there is a dish $r_{j_1}$ with $\sum\limits_{i=1}^Nd^*_{i,j_1}<\left\lfloor\frac{NL}{M}\right\rfloor$ and a dish $r_{j_2}$ with $\sum\limits_{i=1}^Nd^*_{i,j_2}>\left\lceil\frac{NL}{M}\right\rceil$. In such a case, we have $\sum\limits_{i=1}^Nd^*_{i,j_2}>\sum\limits_{i=1}^Nd^*_{i,j_1}+1$, which leads to
\begin{eqnarray}
\sum\limits_\Theta \sum\limits_Q u\Big(q,\sum\limits_{i=1}^Nd^*_{i,j_1}+1\Big)f(q|\theta)p(\theta)>\nonumber\\
\sum\limits_\Theta \sum\limits_Q u\Big(q,\sum\limits_{i=1}^Nd^*_{i,j_2}\Big)f(q|\theta)p(\theta),\label{36}
\end{eqnarray}
From (\ref{36}), we can see that the customer who has requested dish $r_{j_2}$ can obtain higher utility by unilaterally deviating his/her decision to requesting dish $r_{j_1}$. Therefore, $\mathbf D^{*}$ is not a Nash equilibrium of the Indian Buffet Game with budget constraint $L$, and thus we have $\sum\limits_{i=1}^Nd^*_{i,j}=\left\lfloor\frac{NL}{M}\right\rfloor\mbox{ or }\left\lceil\frac{NL}{M}\right\rceil$, when $n_T\ge \left\lceil\frac{NL}{M}\right\rceil$. This completes the proof of the theorem.

\end{IEEEproof}
\section{Non-Bayesian Social Learning}
In the previous two sections, we have analyzed the proposed Indian Buffet Game and characterized the corresponding equilibrium. From the analysis, we can see that the equilibrium highly depends on customers' belief $\mathbf P=\{\mathbf p_j, j=1,2,...,M\}$, i.e., the estimated distribution of the dish state $\bm \theta=\{\theta_j,j=1,2,...,M\}$. The more accurate the belief, the better best response customers can make and thus the better utility customers can obtain. Therefore, it is very important for customers to improve their belief by exploiting from their received signals. In this section, we will discuss the learning process in the proposed Indian Buffet Game. Specifically, we propose an effective non-Bayesian social learning algorithm that can guarantee customers to learn the true system state. Note that since the learning process of different dish state $\theta_j$ are independent of each other, in the rest of this section, we omit the dish index $j$ for notation simplification.

\subsection{Strong Convergence and Weak Convergence}
Suppose the true dish state is $\theta^*$, given customers' belief at time slot $t$, $\mathbf p^{(t)}=\{p^{(t)}(\theta),\forall\ \theta\in\Theta\}$, their belief at time slot $t+1$, $\mathbf p^{(t+1)}=\{p^{(t+1)}(\theta),\forall\ \theta\in\Theta\}$, can be updated by
\begin{equation}
p^{(t+1)}(\theta) = \frac{1}{N}\sum_{i=1}^N\left[d_{i}^{(t+1)}\mu_{i}^{(t+1)}(\theta) + \Big(1-d_{i}^{(t+1)}\Big)p^{(t)}(\theta)\right],\label{NB2}
\end{equation}
where $d^{(t+1)}_i=1$ or $0$ is customer $i$'s decision, and $\mu_{i}^{(t+1)}(\theta)$ is the intermediate belief updated by Bayesian learning rule for customers who have requested the dish and received some signal $s_i^{(t+1)}\sim f(\cdot|\theta^*)$,
\begin{equation}
\mu_{i}^{(t+1)}(\theta)=\frac{f(s_{i}^{(t+1)}|\theta)p^{(t)}(\theta)}{\sum_\Theta {f(s_{i}^{(t+1)}|\theta)p^{(t)}(\theta)}}, \quad \forall\ \theta\in\Theta.\label{38}
\end{equation}
\begin{definition}
A learning rule has the \emph{strong convergence} property if and only if the learning rule can learn the true state in probability as follows:
\begin{equation}
\left\{\begin{array}{c}
p^{(t)}(\theta^*)\rightarrow 1,\\
p^{(t)}(\forall \ \theta\neq\theta^*)\rightarrow 0,
\end{array}\right.\quad \mbox{ as } t\rightarrow\infty.\label{conjecture}
\end{equation}
\end{definition}

By re-organizing some terms, we can re-write the non-Bayesian learning rule in (\ref{NB2}) as
\begin{equation}
p^{(t+1)}(\theta)\! =\! p^{(t)}(\theta)\!+\!\frac{1}{N}\sum_{i=1}^Nd_{i}^{(t+1)}\!\!\left(\frac{f(s_{i}^{(t+1)}|\theta)}{\lambda(s_i^{(t+1)})}-1\right)\!\!p^{(t)}(\theta),\label{NB3}
\end{equation}
with
\begin{equation}
\lambda(s_i^{(t+1)})=\sum\limits_\Theta {f(s_{i}^{(t+1)}|\theta)p^{(t)}(\theta)}.\label{41}
\end{equation}

From (\ref{41}), we can see that $\lambda(s_i^{(t+1)})$ is the estimation of the probability distribution of the signal $s_i^{(t+1)}$ at next time slot. With $\lambda(s_i^{(t+1)})$, we can define a weak convergence, compared with the strong convergence in (\ref{conjecture}), as follows.

\begin{definition}
A learning rule has the \emph{weak convergence} property if and only if the learning rule can learn the true state in probability as follows:
\begin{equation}
\lambda(s)=\sum\limits_\Theta {f(s|\theta)p^{(t)}(\theta)}\rightarrow f(s|\theta^*), \forall\ s\in Q, \ \mbox{as}\ t\rightarrow \infty.\label{weak}
\end{equation}
\end{definition}

Notice that the weak convergence is sufficient for the proposed Indian Buffet Game since the objective of learning here is to find an accurate estimate of the expected utilities of customers and thus derive the true best response. According to (\ref{exutility}), we can see that the signal distribution $\sum_\Theta f_j(q_j|\theta_j)p_j(\theta_j)$ is a sufficient statistic of the expected utility function. Therefore, if we can show that the proposed social learning algorithm have the weak convergence property, then we are able to derive the true best response for customers in the proposed Indian Buffet Game. In the following, we will prove theoretically that the proposed learning algorithm in (\ref{NB2}) indeed has the weak convergence property. We will also show with simulation that the proposed learning algorithm in (\ref{NB2}) have the strong convergence property.

\subsection{Proof of Weak Convergence}
Let us first define a probability triple $(\Omega,\mathcal F, \mathbb P^\theta)$ for some specific dish state $\theta\in\Theta$, where $\Omega$ is the space containing sequences of realizations of the signals $s_i^{(t)}\in Q$, $\mathcal F$ is the $\sigma$-field generated by $\Omega$, i.e., a set of subsets of $\Omega$, and $\mathbb P^\theta$ is the probability measure induced over sample paths in $\Omega$, i.e., $\mathbb P^\theta=\bigotimes^\infty_{t=1}f(\cdot|\theta)$. Moreover, we use $\mathbb E^\theta[\cdot]$ to denote the expectation operator associated with measure $\mathbb P^\theta$, and define $\mathcal F_t$ as the smallest $\sigma$-field generated by the past history of all customers' observations up to time slot $t$. To prove the weak convergence in (\ref{weak}), we start by showing that the belief sequence $\{p^{(t)}(\theta^*)\}$ converges to a positive number as $t\rightarrow\infty$ by the following lemmas.
\begin{lemma}
Suppose the true dish state is $\theta^*$, all customers update their belief $\mathbf p$ according to the non-Bayesian learning rule in (\ref{NB2}) and their prior belief $\mathbf p^{(0)}$ satisfies $p^{(0)}(\theta^*)>0$, then, the belief sequence $\{p^{(t)}(\theta^*)\}$ converges to a positive number as $t\rightarrow\infty$.
\end{lemma}
\begin{IEEEproof}
From (\ref{NB2}) and (\ref{38}), we can see that if $p^{(t)}(\theta)>0$, then $p^{(t+1)}(\theta)>0$. Since the prior belief satisfies $p^{(0)}(\theta^*)>0$, according to the method of induction, we have the belief sequence $\{p^{(t)}(\theta^*)\}>0$.

According to (\ref{NB3}), for the true dish state $\theta^*$, we have
\begin{equation}
p^{(t+1)}(\theta^*)\! \!= \!p^{(t)}(\theta^*)\!+\!\frac{1}{N}\sum_{i=1}^Nd_{i}^{(t+1)}\!\!\left(\frac{f(s_{i}^{(t+1)}|\theta^*)}{\lambda(s_i^{(t+1)})}-1\right)\!\!p^{(t)}(\theta^*).\label{qaz}
\end{equation}
By taking expectation over $\mathcal F_t$ on both sides of (\ref{qaz}), we have
\begin{align}
&\!\!\mathbb E^{\theta^*}\left[p^{(t+1)}(\theta^*)|\mathcal F_t\right] = p^{(t)}(\theta^*)\nonumber+\\
& \frac{1}{N}\sum_{i=1}^N\mathbb E^{\theta^*}\left[d_{i}^{(t+1)}\left(\frac{f(s_{i}^{(t+1)}|{\theta^*})}{\lambda(s_i^{(t+1)})}-1\right)\bigg|\mathcal F_t\right]p^{(t)}({\theta^*}).\label{wsx}
\end{align}
Since customer's decision $d^{(t+1)}_i$ and the signal he/she receives $s^{(t+1)}_i$ are independent of each other over $\mathcal F_t$, we can separate the expectation in the second term of (\ref{wsx}) as
\begin{align}
&\mathbb E^{\theta^*}\left[d_{i}^{(t+1)}\cdot\left(\frac{f(s_{i}^{(t+1)}|{\theta^*})}{\lambda(s_i^{(t+1)})}-1\right)\bigg|\mathcal F_t\right]=\nonumber\\
&\quad\mathbb E^{\theta^*}\left[d_{i}^{(t+1)}\big|\mathcal F_t\right]\cdot\mathbb E^{\theta^*}\left[\left(\frac{f(s_{i}^{(t+1)}|{\theta^*})}{\lambda(s_i^{(t+1)})}-1\right)\bigg|\mathcal F_t\right].\label{mlp}
\end{align}
In (\ref{mlp}), $\mathbb E^\theta\left[d_{i}^{(t+1)}\big|\mathcal F_t\right]\ge 0$ since $d_{i}^{(t+1)}$ can only be $1$ or $0$. Moreover, since $g (x) = 1/x$ is a convex function, according to Jensen's inequality, we have
\begin{align}
&\mathbb E^{\theta^*}\left[\frac{f(s_{i}^{(t+1)}|{\theta^*})}{\lambda(s_i^{(t+1)})}\bigg|\mathcal F_t\right]\ge\left(\mathbb E^{\theta^*}\left[\frac{\lambda(s_i^{(t+1)})}{f(s_{i}^{(t+1)}|{\theta^*})}\bigg|\mathcal F_t\right]\right)^{-1}\nonumber\\
&\quad\quad\quad=\left(\sum\limits_Q\frac{\lambda(s_i^{(t+1)})}{f(s_{i}^{(t+1)}|{\theta^*})}f(s_{i}^{(t+1)}|{\theta^*})\right)^{-1}=1.
\end{align}
In such a case, the equation in (\ref{mlp}) is non-negative, which means that in (\ref{wsx}),
\begin{equation}
\mathbb E^{\theta^*}\left[p^{(t+1)}(\theta^*)|\mathcal F_t\right] \ge p^{(t)}(\theta^*).
\end{equation}
Since customers' belief $p^{(t)}(\theta^*)$ is bounded within interval [0,1], according to the martingale convergence theorem \cite{martingale}, we can conclude that the belief sequence $\{p^{(t)}(\theta^*)\}$ converges to a positive number as $t\rightarrow\infty$.
\end{IEEEproof}
\begin{theorem}
In an Indian Buffet restaurant, suppose that the true dish state is $\theta^*$, all customers update their belief $\mathbf p$ using (\ref{NB2}) and their prior belief $\mathbf p^{(0)}$ satisfies $p^{(0)}(\theta^*)>0$, then, the belief sequence $\{p^{(t)}(\theta)\}$ ensures a weak convergence, i.e., for $\forall \ s\in Q$,
\begin{equation}
\lambda(s)=\sum\limits_\Theta {f(s|\theta)p^{(t)}(\theta)}\rightarrow f(s|\theta^*), \quad \mbox{as}\ t\rightarrow \infty.
\end{equation}
\end{theorem}
\begin{IEEEproof}
Let $\mathcal N^{(t+1)}$ denote the set of customers who request the dish at time slot $t+1$. In such a case, we can re-write (\ref{qaz}) as
\begin{equation}
p^{(t+1)}(\theta^*)= \frac{1}{|\mathcal N^{(t+1)}|}\sum_{i\in\mathcal N^{(t+1)}}\frac{f(s_{i}^{(t+1)}|\theta^*)}{\lambda(s_i^{(t+1)})}p^{(t)}(\theta^*),\label{nko}
\end{equation}
where $|\cdot|$ means the cardinality. By taking logarithmic operation on both sides of (\ref{nko}) and utilizing the concavity of the logarithm function, we have
\begin{equation}
\log p^{(t+1)}(\theta^*)\!\ge\!\log p^{(t)}(\theta^*)+ \frac{1}{|\mathcal  N^{(t+1)}|}\!\!\!\sum_{i\in\mathcal N^{(t+1)}}\!\!\!\!\!\!\log \frac{f(s_{i}^{(t+1)}|\theta^*)}{\lambda(s_i^{(t+1)})}.\label{nko2}
\end{equation}
Then, by taking expectation over $\mathcal F_t$ on both sides of (\ref{nko2}), we have
\begin{align}
&\mathbb E^{\theta^*}\left[\log p^{(t+1)}(\theta^*)|\mathcal F_t\right]-\log p^{(t)}(\theta^*)\nonumber\\
&\quad\quad\ge\frac{1}{|\mathcal  N^{(t+1)}|}\sum_{i\in\mathcal N^{(t+1)}}\mathbb E^{\theta^*}\left[\log \frac{f(s_{i}^{(t+1)}|\theta^*)}{\lambda(s_i^{(t+1)})}\bigg|\mathcal F_{t}\right].\label{54}
\end{align}
As to the left hand of (\ref{54}), according to \emph{Lemma 2}, we know that $p^{(t)}(\theta^*)$ will converge as $t\rightarrow \infty$, and thus
\begin{equation}
\mathbb E^{\theta^*}\left[\log p^{(t+1)}(\theta^*)|\mathcal F_t\right]-\log p^{(t)}(\theta^*)\rightarrow 0.\label{55}
\end{equation}
As to the right hand of (\ref{54}), it follows
\begin{align}
\mathbb E^{\theta^*}\left[\log \frac{f(s_{i}^{(t+1)}|\theta^*)}{\lambda(s_i^{(t+1)})}\bigg|\mathcal F_{t}\right]&=-\mathbb E^{\theta^*}\left[\log \frac{\lambda(s_i^{(t+1)})}{f(s_{i}^{(t+1)}|\theta^*)}\bigg|\mathcal F_{t}\right]\nonumber\\
&\ge -\log\mathbb E^{\theta^*}\left[ \frac{\lambda(s_i^{(t+1)})}{f(s_{i}^{(t+1)}|\theta^*)}\bigg|\mathcal F_{t}\right]\nonumber\\
&=0.\label{56}
\end{align}
In such a case, combining (\ref{55}) and (\ref{56}), as $t\rightarrow \infty$, we have
\begin{align}
0 \ge\frac{1}{|\mathcal  N^{(t+1)}|}\sum_{i\in\mathcal N^{(t+1)}}\mathbb E^{\theta^*}\left[\log \frac{f(s_{i}^{(t+1)}|\theta^*)}{\lambda(s_i^{(t+1)})}\bigg|\mathcal F_{t}\right]\ge 0.
\end{align}
By squeeze theorem \cite{sqeeze}, we have for $\forall \ i\in\mathcal N^{(t+1)}$, as $t\rightarrow \infty$,
\begin{align}
&\!\!\!\!\!\!\!\!\!\!\!\!\!\!\!\mathbb E^{\theta^*}\left[\log \frac{f(s_{i}^{(t+1)}|\theta^*)}{\lambda(s_i^{(t+1)})}\bigg|\mathcal F_{t}\right]=\nonumber\\
&\sum\limits_\Theta f(s_{i}^{(t+1)}|{\theta^*})\log \frac{f(s_{i}^{(t+1)}|\theta^*)}{\lambda(s_i^{(t+1)})}\rightarrow 0,\label{finish}
\end{align}
According to Gibbs' inequality \cite{gibbs}, the (\ref{finish}) converges to $0$ if and only if as $t\rightarrow \infty$,
\begin{equation}
\lambda(s_i^{(t+1)})\rightarrow f(s_{i}^{(t+1)}|\theta^*), \quad \forall\ s_{i}^{(t+1)}\in Q.
\end{equation}
This completes the proof of the theorem.
\end{IEEEproof}

\section{Simulation Results}
In this section, we conduct simulations to verify the performance of the proposed non-Bayesian social learning rule and recursive best response algorithms. We simulate an Indian Buffet restaurant with five dishes $\{r_1,r_2,r_3,r_4,r_5\}$ and five possible dish states $\theta_j\in\{1,2,3,4,5\}$. Each dish is randomly assigned with a state. After requesting some specific dish $r_j$, customer $i$ can infer the quality of the dish and receive a signal $s_{i,j}\in\{1,2,3,4,5\}$ obeying the conditional distribution that
\begin{equation}
f_j(s_{i,j}|\theta_j)=\left\{
\begin{array}{ll}
w, & \quad\mbox{if }s_{i,j}=\theta_j;\\
(1-w)/4, & \quad\mbox{if }s_{i,j}\neq\theta_j.
\end{array}\label{simu1}
\right.
\end{equation}
The parameter $w$ can be interpreted as the quality of the signal or customers' detection probability. When the signal quality $w$ is close to $1$, the customers' received signal is more likely to reflect the true dish state. Note that $w$ should satisfy $w\ge 1/5$; otherwise, the true state can never be learned correctly. With the signals, customers can update their belief $\mathbf P$ cooperatively at the next time slot and then make their decisions sequentially. Once the $i$-th customer makes the dishes selection, he/she reveals his/her decisions to other customers. After all customers make their decisions, they begin to share the corresponding dishes they have requested. The customer $i$'s utility of requesting dish $r_j$ is given by
\begin{equation}
u_{i,j}=\gamma_i\frac{s_{i,j}R}{N_j}-c_j,\label{simu}
\end{equation}
where $\gamma_i$ is a utility coefficient for customer $i$ since different customers may have different utilities regarding same reward, $s_{i,j}$ is a realization of dish quality, as well as the signal inferred by customer $i$, $R$ is the basis award of requesting each dish as $R=10$, $N_j$ is the overall number of customers requesting dish $r_j$ and $c_j$ is the cost of requesting dish $r_j$ as $\{c_j=1, \forall j\}$. From (\ref{simu1}) and (\ref{simu}), we can see that by requesting dish with higher level of state, e.g., $\theta_j=5$, customers can obtain higher utilities. However, customers are unknown about the dish state and have to estimate it through social learning. On the other hand, we can also see that the more customers requesting the same dish, the less utility each customer can obtain, which embodies the negative network externality.

\subsection{Indian Buffet Game without Budget Constraint}

In this subsection, we evaluate the performance of the proposed best response algorithm for Indian Buffet Game without budget constraint. We first simulate the homogenous case to verify the threshold property of the Nash equilibrium matrix, i.e., \emph{Theorem 2}, and the influence of different decision making orders on customers' utilities, i.e., making decisions earlier may have advantage. Then, we compare the performance of the proposed best response algorithm, i.e., Algorithm 1, with the performances of other algorithms under heterogenous settings.

For the homogenous case, we set all customers' utility coefficients as $\gamma_i=1$. The customers' prior belief regarding the dish state starts with a uniform distribution, i.e., $\{p^{(0)}_j(\theta)=0.2, \forall j,\theta\}$. The dish state is set as $\mathbf \Theta=[1,2,3,4,5]$, i.e., $\theta_j=j$, in order to verify different threshold structures for different dish states as illustrated in \emph{Theorem 2}. At each time slot, we let customers sequentially make decisions according to Algorithm 1 and then update their belief according to the non-Bayesian learning rule. The game is played time slot by time slot until customers' belief $\mathbf P^{(t)}$ converges. In the first simulation, we set the number of customers as $N=10$ to specifically verify the threshold structure of Nash equilibrium matrix. Table~\ref{brm} shows the Nash equilibrium matrix $\mathbf D^*$ derived by Algorithm 1 after customers' belief $\mathbf P^{(t)}$ converges, where each column contains one customer's decisions $\{d_{i,j}, \forall j\}$ and each row contains all customers' decisions on one specific dish $r_j$, i.e., $\{d_{i,j}, \forall i\}$. From the table, we can see that once some customer does not request some specific dish, all the subsequent customers will not request that dish, which is consistent with the conclusion in \emph{Theorem 2}. Moreover, since requesting the dish with higher level of state, e.g., $\theta_5=5$, can obtain higher utilities, we can see that most customers decided to request dish $r_5$.
\begin{table}[!t]
    \caption{Nash equilibrium matrix $\mathbf D^*$}
    \begin{center}
    \normalsize
    \begin{tabular}{c|ccccccccccc}
    & &  \emph{1}  &   \emph{2}  &   \emph{3}  &   \emph{4}  &   \emph{5}  &   \emph{6}  &   \emph{7}  &   \emph{8}  &   \emph{9}  &   \emph{10}\\\hline\\
    $r_1$ & &  1  &   1  &   0  &   0  &   0  &   0  &   0  &   0  &   0  &   0\\
    $r_2$ & &  1  &   1  &   1  &   1  &   0  &   0  &   0  &   0  &   0  &   0\\
    $r_3$ & &  1  &   1  &   1  &   1  &   1  &   0  &   0  &   0  &   0  &   0\\
    $r_4$ & &  1  &   1  &   1  &   1  &   1  &   1  &   0  &   0  &   0  &   0\\
    $r_5$ & &  1  &   1  &   1  &   1  &   1  &   1  &   1  &   1  &   0  &   0
    \end{tabular}
    \end{center}\label{brm}
\end{table}
\begin{figure}[!t]
\centering
\centerline{\epsfig{figure=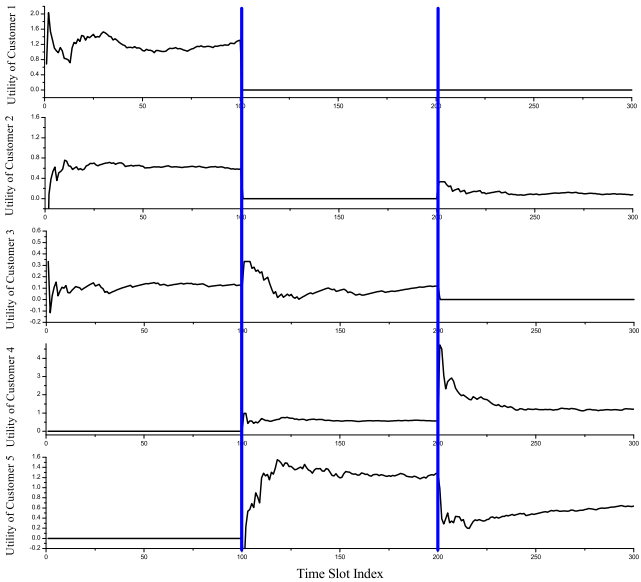,width=8.5cm}}
\caption{Each customer's utility in homogenous case without budget constraint.} \label{noeach}
\end{figure}
From Table~\ref{brm}, we can see that customers who make decisions earlier have advantage, e.g., customer 1 can request all dishes while customer 8 can only request one dish. Therefore, in the second simulation of the homogenous case, we dynamically adjust the order of decision making to ensure the fairness. In this simulation, we assume that there are 5 customers with a common utility coefficient $\gamma_i=0.4$. In Fig.\,\ref{noeach}, we show all customers' utilities along with the simulation time, where the order of decision making changes every 100 time slots. In the first 100 time slots, where the order of decision making is $1\rightarrow2\rightarrow3\rightarrow4\rightarrow5$, we can see that customer 1 obtains the highest utility and customer 4 and 5 receive 0 utility since they have not requested any dish. In the second 100 time slots, we reverse the decision making order as $5\rightarrow4\rightarrow3\rightarrow2\rightarrow1$, which leads to that customer 1 and 2 receive 0 utility. Therefore, by periodically changing the order of decision making, we can expect that the expected utilities of all customers will be the same after a period of time.

\begin{figure}[!t]
\centering
\centerline{\epsfig{figure=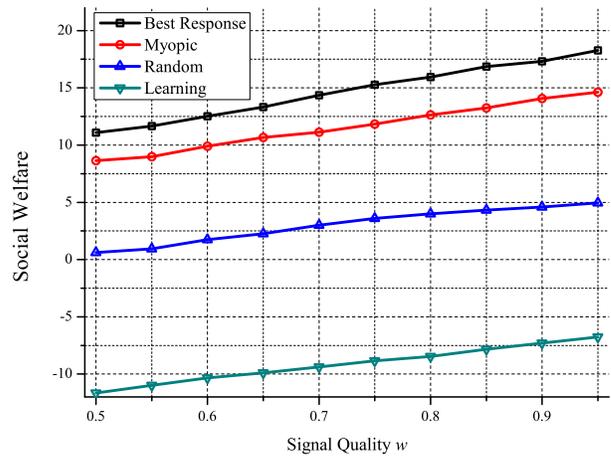,width=8.0cm}}
\caption{Social welfare comparison without budget constraint.} \label{nosocial}
\end{figure}
For the heterogenous case, we randomize each customer's utility coefficient $\gamma_i$ between 0 and 1 and set their prior belief as $\{p^{(0)}_j(\theta)=0.2, \forall j,\theta\}$. In this simulation, we compare the performance in terms of customers' social welfare, which is defined as the total utilities of all customers, among different kinds of algorithms listed as follows:
\begin{itemize}
    \item \textbf{Best Response}: The proposed recursive best response algorithm in Algorithm 1 with non-Bayesian learning.
    \item \textbf{Myopic}: At each time slot, customer $i$ requests dishes according to current observation $\mathbf n_{i}=\{n_{i,j},\forall j\}$ without social learning.
    \item \textbf{Learning}: At each time slot, each customer requests dishes purely based on the updated belief $\mathbf P^{(t)}$ using non-Bayesian learning rule without considering the negative network externality.
    \item \textbf{Random}: Each customer randomly requests dishes.
\end{itemize}
For the myopic and learning strategies, customer $i$'s expected utility of requesting dish $r_j$ can be calculated by
\begin{align}
U^{\mbox{m}}_{i,j}& = \sum_\Theta  {\sum_Q {u_{i,j}({q_j},{n_{i,j}} + {d_{i,j}})f_j ({q_j}|{\theta _j}){p^{(0)}_j}({\theta _j})} },\\
U^{\mbox{l}}_{i,j}& = \sum_\Theta  {\sum_Q {u_{i,j}({q_j},{d_{i,j}})f_j ({q_j}|{\theta _j}){p^{(t)}_j}({\theta _j})} }.
\end{align}
With these expected utilities, both myopic and learning algorithm can be derived by (\ref{bestR1}). We can see that the myopic strategy does not consider social learning while the learning strategy does not involve negative network externality. In the simulation, we average these four algorithms over hundreds of realizations. Fig.\,\ref{nosocial} shows the performance comparison result, where the x-axis is the signal quality $w$ varying from $0.5$ to $0.95$ and y-axis is the social welfare averaged over hundred of time slots. From the figure, we can see with the increase of signal quality, the social welfare keeps increasing for all algorithms. Moreover, we can also see that our best response algorithm performs the best while the learning algorithm performs the worst. This is because, with the learning algorithm, customers can gradually learn the true dish states and then request the dish without considering other customers' decisions. In such a case, too many customers may request the same dishes and each customer's utility is dramatically decreased due to the negative network externality. For the myopic algorithm, although customers can not learn the true dish states, by considering other customers' decisions, each customer can avoid requesting dishes which have been over-requested. Therefore, we can conclude that our proposed best response algorithm achieve the best performance through considering the negative network externality and using social learning to estimate the dish state.

\subsection{Indian Buffet Game with Budget Constraint}

\begin{figure}[!t]
\centering
\centerline{\epsfig{figure=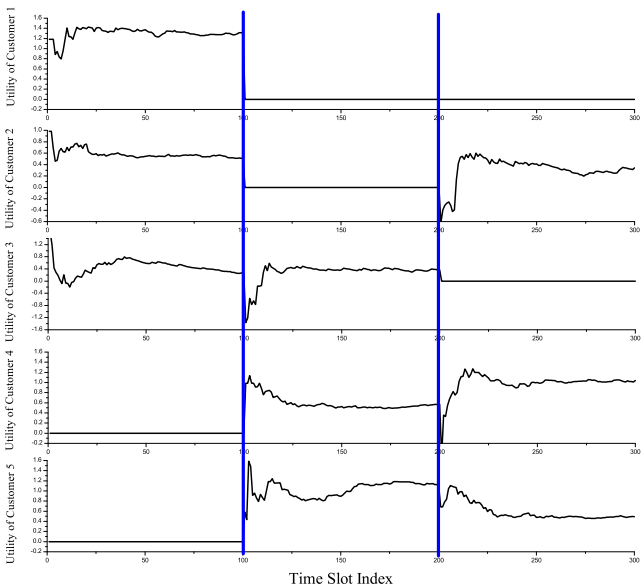,width=8.5cm}}
\caption{Each customer's utility in homogenous case with budget constraint.} \label{witheach}
\end{figure}

In this subsection, we evaluate the performance of the proposed best response algorithm for Indian Buffet Game with budget constraint $L=3$. Similar to the previous subsection, we start from the homogenous case, where all customers' utility coefficients are set as $\gamma_i=1$. In the first simulation, we set all dish states as $\theta_j=5$ to verify the property of the Nash equilibrium matrix illustrated in \emph{Theorem 4}. Table~\ref{brm2} shows the Nash equilibrium matrix $\mathbf D^*$ derived by Algorithm 2. We can see that each dish has been requested by $N*L/M=10*3/5=6$ customers, which is consistent with the conclusion in \emph{Theorem 4}. In the second simulation, we dynamically change the order of customers' sequential decision making and illustrate each customer's utility along with simulation time in Fig.\,\ref{witheach}, from which we can see similar phenomenons as Indian Buffet Game without budget constraint.
\begin{table}[!t]
    \caption{Nash equilibrium matrix $\mathbf D^*$}
    \begin{center}
    \normalsize
    \begin{tabular}{c|ccccccccccc}
    & &  \emph{1}  &   \emph{2}  &   \emph{3}  &   \emph{4}  &   \emph{5}  &   \emph{6}  &   \emph{7}  &   \emph{8}  &   \emph{9}  &   \emph{10}\\\hline\\
    $r_1$ & &  1  &   1  &   1  &   1  &   1  &   1  &   0  &   0  &   0  &   0\\
    $r_2$ & &  1  &   1  &   1  &   1  &   0  &   0  &   1  &   1  &   0  &   0\\
    $r_3$ & &  1  &   1  &   1  &   1  &   0  &   0  &   0  &   0  &   1  &   1\\
    $r_4$ & &  0  &   0  &   0  &   0  &   1  &   1  &   1  &   1  &   1  &   1\\
    $r_5$ & &  0  &   0  &   0  &   0  &   1  &   1  &   1  &   1  &   1  &   1
    \end{tabular}
    \end{center}\label{brm2}
\end{table}

\begin{figure}[!t]
\centering
\centerline{\epsfig{figure=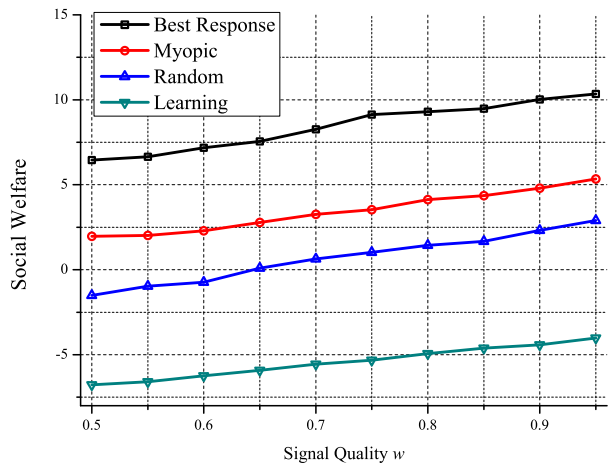,width=8.0cm}}
\caption{Social welfare comparison with budget constraint.} \label{withsocial}
\end{figure}

For the heterogenous case, we randomize each customer's utility coefficient $\gamma_i$ within $[0,1]$ and compare the performance of our proposed best response algorithm, i.e., Algorithm 2, with myopic, learning and random algorithms in terms of customers' social welfare. For the myopic, learning and random algorithms, same budget constraint is adopted, i.e., each customer can at most request $3$ dishes. Fig.\,\ref{withsocial} shows the performance comparison result, from which we can see that our best response algorithm performs the best while the learning algorithm performs the worst.

\begin{figure}[!t]
\centering
\centerline{\epsfig{figure=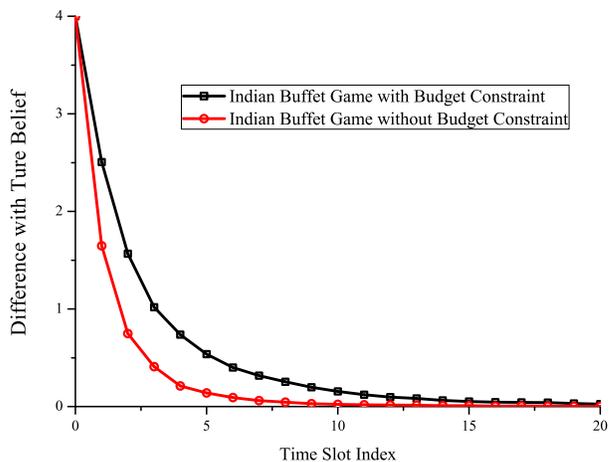,width=8.0cm}}
\caption{Performance of the non-Bayesian social learning rule.} \label{learnp}
\end{figure}

\subsection{Non-Bayesian Social Learning Performance}

In this subsection, we evaluate the performance of the proposed non-Bayesian social learning rule. At the beginning of the simulation, we randomize the states of 5 dishes and assign customers' prior belief regarding each dish state with uniform distribution, i.e., $\{p_j(\theta=0.2), \forall j, \theta\}$. After requesting the chosen dishes, each customer can receive signals following the conditional distribution defined in (\ref{simu1}) with signal quality $w=0.6$. Fig.\,\ref{learnp} shows the learning curve of the Indian Buffet Game without and without budget constraint, respectively. The y-axis is the difference between customers' belief at each time slot $\mathbf P^{(t)}$ and the true belief $\mathbf P^{o}$, which can be calculated by $||\mathbf P^{(t)}-\mathbf P^{o}||_2$. From the figure, we can see that customers can learn the true dish states within $15$ time slots. Moreover, the convergence rate of the case without budget constraint is faster than that of the case with budget constraint. This is because, due to the budget constraint, each customer requests fewer dishes at each time slot and thus receives fewer signals regarding the dish state, which will inevitably slow down the customers' learning speed.

\section{Conclusion}

In this paper, we we proposed a general framework, called Indian Buffet Game, to study how users make multiple concurrent selections under uncertain system state. We studied the game under two different scenarios: customers request multiple dishes without budget constraint and with budget constraint, respectively. We designed best response algorithms for both cases to find the subgame perfect Nash equilibrium, and discussed the simplified homogeneous cases to better understand the proposed Indian Buffet Game. We also designed a non-Bayesian social learning rule for customers to learn the dish state and theoretically prove its convergence. Simulation results show that our proposed algorithms achieve much better performance than myopic, learning and random algorithms. Moreover, the proposed non-Bayesian learning algorithm can help customers learn the true system state with a fast convergence rate.
\bibliographystyle{IEEEtran}
\bibliography{list}

\end{document}